\documentclass[preprint]{aastex}
\usepackage{psfig}
\begin{document}

\def\l{\hat{l}}
\def\e{\epsilon}
\def\f{{\rm f}}
\def\ne{n_{\rm e}}
\def\sT{\sigma_{\rm T}}

\title{Compton Scattering in Static and Moving Media.
  \\ II. System-Frame Solutions for Spherically Symmetric Flows}
\author{Dimitrios Psaltis}
\affil{Harvard-Smithsonian Center for Astrophysics,
       60 Garden St., Cambridge, MA 02138\\
       and\\ 
       Center for Space Research, 
       Massachusetts Institute of Technology
       Cambridge, MA 02139}
\email{demetris@space.mit.edu}

\slugcomment{Submitted to {\em The Astrophysical Journal}.}

\begin{abstract}
  I study the formation of Comptonization spectra in spherically
  symmetric, fast moving media in a flat spacetime. I analyze the
  mathematical character of the moments of the transfer equation in
  the system-frame and describe a numerical method that provides fast
  solutions of the time-independent radiative transfer problem that
  are accurate in both the diffusion and free-streaming regimes. I
  show that even if the flows are mildly relativistic ($V\sim 0.1$,
  where $V$ is the electron bulk velocity in units of the speed of
  light), terms that are second-order in $V$ alter the emerging
  spectrum both quantitatively and qualitatively. In particular, terms
  that are second-order in $V$ produce power-law spectral tails, which
  are the dominant feature at high energies, and therefore cannot be
  neglected. I further show that photons from a static source are
  upscattered by the bulk motion of the medium even if the velocity
  field does not converge. Finally, I discuss these results in the
  context of radial accretion onto and outflows from compact objects.
\end{abstract}


\section{INTRODUCTION}

Compton scattering in quasi-spherical accretion flows is thought to be
responsible for the X-ray spectra of many accreting compact objects.
For example, the hard X-ray spectra of weakly-magnetic accreting
neutron stars can be accounted for if a geometrically thick, hot
medium surrounds the neutron star and inner accretion disk (see, e.g.,
Psaltis, Lamb, \& Miller 1995). The spectral signature of isolated
neutron stars accreting from the interstellar medium, which is
required for their positive identification, is determined mainly by
the efficiency of Compton upscattering of thermal photons emitted from
the stellar surfaces (see, e.g., Zampieri et al.\ 1995). Moreover,
Compton scattering plays a major role in the formation of the
high-energy spectra of advection-dominated accretion flows, which may
be the relevant mode of accretion onto galactic and supermassive black
holes (see, e.g., Esin et al.\ 1997).

Compton scattering in static and moving media, even in plane-parallel
or spherical symmetry, couples strongly all three independent
phase-space coordinates of the photons, namely their energy, direction
of propagation, and spatial coordinate. For this reason, Monte Carlo
methods have been used extensively in calculating Comptonization
spectra, because of their easy implementation and natural treatment of
the diverse length scales in an accretion flow (see, e.g., Laurent \&
Titarchuk 1999 for a recent study; methods and results based on Monte
Carlo treatments have been reviewed by Pozdnyakov, Sobol, \& Sunyaev
1983 and will not be discussed here).  Alternatively, Compton
scattering can be studied by solving the energy-dependent radiative
transfer equation or its moments. Implementing such a method is more
difficult and often relies on a number of approximations for
describing the interaction of radiation with matter (see, however,
Poutanen \& Svensson 1996 and Hsu \& Blaes 1998 for complete solutions
in static media). However, at the same time it is substantially faster
and offers significant advantages when the radiative transfer problem
is coupled to a solution of the fluid dynamics, as well as in treating
steep spectra and processes that do not conserve photons.

Over the last twenty five years several authors have derived the
photon kinetic or radiative transfer equation and their moments for
Compton scattering in static and moving media (see, e.g., Kompaneets
1957; Babuel-Peyrissac \& Rouvillois 1969; Pomraning 1973; Chan \&
Jones 1975; Payne 1980; Blandford \& Payne 1981a; Thorne 1981; Fukue,
Kato, \& Matsumoto 1985; Madej 1989, 1991; Titarchuk 1994). The moment
equations derived by these various sets of authors have been widely
used in studies of Comptonization by static media (see, e.g., Katz
1976; Shapiro, Lightman, \& Eardly 1976; Sunyaev \& Titarchuk 1980) or
by strong shocks and accretion flows onto compact objects (see, e.g.,
Blandford \& Payne 1981b; Payne \& Blandford 1981; Lyubarskij \&
Sunyaev 1982; Colpi 1988; Riffert 1988; Mastichiadis \& Kylafis 1992;
Titarchuk \& Lyubarskij 1995; Turolla et al. 1996; Titarchuk,
Mastichiadis, \& Kylafis 1997), either under the diffusion
approximation or with closure relations that were specified a priori.
Most of the above analyses of Compton scattering in moving media were
performed in the system frame and to first order in the electron bulk
velocity. On the other hand, Schmid-Burgk (1978), Zane et al.\ (1996),
and Titarchuk \& Zannias (1998) have solved the general-relativistic
radiative transfer equation for Compton scattering without the need
for specifying a priori closure relations.

In Paper I (Psaltis \& Lamb 1997) of this series, the time-dependent
photon kinetic and radiative transfer equations were derived, as well
as their zeroth and first moments that describe absorption, emission,
and spontaneous and induced Compton scattering in static and moving
media, correcting various errors in the literature. The system-frame
equations that were derived are valid to first order in $\epsilon/m_e$
and $T_e/m_e$, and to {\it second} order in $V$, where $m_e$ and $T_e$
are the electron mass and temperature, $\epsilon$ is the photon
energy, and $V$ is the bulk velocity of the electrons in units of the
speed of light; the fluid-frame equations that were derived are valid
for arbitrary values of the bulk velocity $V$. Using these equations
it was argued in Paper I that the effects of Comptonization by the
bulk electron velocity that are described by the terms that are
second-order in $V$ are qualitatively different than and can become
at least as important as the effects described by the terms that are
first-order in $V$, even when $V$ is small.  As a result, these
second-order terms should generally be retained (see also Yin \&
Miller 1995).

In this paper, I describe a numerical algorithm for the solution of
time-independent radiative transfer problems in systems with spherical
symmetry.  Even though the natural reference frame for solving
transport problems is the one comoving with the flow, I work in the
system frame, which can be any inertial frame in which the central
object is at rest. This is important for the fast algorithm described
below, because the moment equations in the system frame are always
parabolic (as shown in \S3) in contrast to the fluid-frame equations
(Turolla, Zampieri, \& Nobili 1995; Smit, Cernohorsky, \& Dullemond
1997; see also K\"orner \& Janka 1992). Moreover, in problems of mass
accretion onto a compact star, the boundary conditions are often more
easily implemented in the system frame.  I also neglect any general
relativistic effects, which can be shown to introduce only small
quantitative corrections to the emerging spectra (Papathanassiou \&
Psaltis 2000), and truncate the transfer equation keeping only terms
up to second order in $V$.

The algorithm used is a generalization of the method of variable
Eddington factors (Mihalas 1980; see also Mihalas 1978, p.~157--158).
It is therefore accurate for systems of arbitrary optical depth and
does not require a priori specification of closure relations. It is
based on the iterative solution of both the transfer equation and the
systems of its first two moments and hence the validity and accuracy
of the solution can be verified explicitly at the end of each run. The
method also requires only a small number of iterations in order to
converge to the solution and is therefore ideal for future extensions
to time-dependent and multi-dimensional transport problems.

In \S2, I summarize the results from Paper I and introduce the
notation. In \S3, I discuss the mathematical character of the system
of equations and in \S4, I present the numerical algorithm. In \S5, I
describe the results of numerical solutions of the transfer equations
for a variety of situation. Finally, in \S6, I briefly summarize the
implications of the above for Comptonization spectra in static and
moving media.

\section{THE ELECTRON GAS AND RADIATION FIELD}

{\it Units}---Throughout this paper I set $h=k_{\rm B}=c=1$, where
$h$ is Planck's constant, $k_{\rm B}$ is Boltzmann's constant, and $c$
is the speed of light. I also normalize all the spatial coordinates
to the inner radius of the flow, the electron temperature $T_e$ and
photon energy $\epsilon$ to the electron rest mass $m_e$, and the
electron density $n_e$ to its value at the inner radius of the flow.
Finally, I normalize the emission and absorption coefficients
$\eta(r,\epsilon)$ and $\chi(r,\epsilon)$ to the inverse of the
electron scattering mean-free path in the Thomson limit.

{\it The electron gas}---Let $\vec{u}$ be the system-frame
three-velocity of a given electron.  I define the local bulk velocity
$\vec{V}$ of the electrons measured in the system frame as
\begin{equation}
 \vec{V}\equiv\langle \vec{u}\rangle\;,
\end{equation}
where the sharp brackets indicate the average over the local electron
velocity distribution. I also assume that the electron velocity
distribution in the fluid frame is a relativistic Maxwellian and
therefore
\begin{equation}
 \langle u^2 \rangle \simeq V^2+ 3 T_e \;,
  \label{u2ave}
 \end{equation}
where I have neglected terms of order $V^2 T_e$ and higher.

{\it The radiation field}---I describe the radiation field at any
position in the system frame with coordinate vector $\vec{r}$ using
the monochromatic specific intensity $I(\vec{r},\hat{l},\epsilon)$,
where $\hat{l}$ is the direction of propagation and $\epsilon$ is the
photon energy. Here I consider only unpolarized radiation and hence
I have suppressed the dependence of the specific intensity on
polarization. Because of the assumed spherical symmetry of the problem,
the radiation field depends only on the radial distance from the
center of symmetry (i.e., the $r$--component of the vector $\vec{r}$),
on the angle $\theta$ between the radial direction and the direction
of propagation (i.e., $\cos\theta=\hat{l}\cdot\vec{r}/r$), and on the
photon energy $\epsilon$.

I define the first five moments of the monochromatic specific
intensity as
 \begin{eqnarray}
 J(\vec{r},\epsilon) & \equiv & \frac{1}{4\pi}
  \int_\Omega I(\vec{r},\hat{l},\epsilon) d\Omega \\
 H^{i}(\vec{r},\epsilon) & \equiv & \frac{1}{4\pi}
  \int_\Omega I(\vec{r},\hat{l},\epsilon) l^i d\Omega \\
 K^{ij}(\vec{r},\epsilon) & \equiv &  \frac{1}{4\pi}
  \int_\Omega I(\vec{r},\hat{l},\epsilon) l^i l^j d\Omega
\\
 Q^{ijk}(\vec{r},\epsilon) & \equiv &  \frac{1}{4\pi}
  \int_\Omega I(\vec{r},\hat{l},\epsilon) l^il^jl^k
d\Omega \\
 R^{ijkl}(\vec{r},\epsilon) & \equiv & \frac{1}{4\pi}
  \int_\Omega I(\vec{r},\hat{l},\epsilon)
l^il^jl^kl^ld\Omega\;.
 \end{eqnarray}
The non-zero components of the first five moments of the
specific intensity for systems with spherical symmetry are
given in Appendix~A.

{\it The transfer equation and its moments}---Following Mihalas (1980;
see also Mihalas 1978), I solve the radiative transfer equation in a
plane that contains the center of symmetry of the system, between the
inner and outer radii $r_{\rm in}$ and $r_{\rm out}$ of the flow,
using the coordinates $z\equiv r\cos\theta$ and $p\equiv r\sin\theta$.

The radiative transfer equation along a ray of constant
impact parameter $p$ is 
\begin{equation}
\pm \frac{\partial}{\partial \tau_z} I^\pm(z,p,\epsilon) = 
   (1-2\epsilon)I^\pm(z,p,\epsilon) -
S^\pm(z,p,\epsilon)\;,
   \label{RTEq}
\end{equation}
where the signs `$+$' and '$-$' correspond to the equations for the
outgoing and incoming rays respectively and I have neglected the
effects of induced Compton scattering. In equation~(\ref{RTEq}) I
have used the optical depth along the ray defined as $d\tau_z\equiv
-n_e \sigma_T [1+\chi(r,z,\epsilon)] dz$, where $\tau_z=0$ at the
outer boundary and $\sigma_T$ is the angle-integrated Thomson
scattering cross section.  $S^\pm(z,p,\epsilon)$ is the generalized
source function for absorption, emission, and Compton scattering in
systems with spherical symmetry and is given in Appendix~B.

The equations for the moments of the specific intensity can be derived
either directly from equation (\ref{RTEq}) or from equations (34) and
(40) of Paper I.  Written in compact form and using the dimensionless
quantities defined above, the moment equations are\footnote{Note that
the $B_3$ term reported in Paper I (Psaltis \& Lamb 1997) contained
a typo that has been corrected here.}
\begin{eqnarray}
A_1 J + A_2 \epsilon \partial_\epsilon J + A_3
\epsilon^2\partial^2_\epsilon J +
   A_4 H^r + A_5 \epsilon \partial_\epsilon H^r +
\partial_{\tau_r}H^r
   & = & C_1 \label{0th3}\\
B_1 J + B_2 \epsilon \partial_\epsilon J +
f^{rr}\partial_{\tau_r}J+
   B_3 H^r + B_4 \epsilon \partial_\epsilon H^r +  B_5
\epsilon^2\partial^2_\epsilon H^r
   & = & C_2 \label{1st}\;,
\end{eqnarray}
where $\partial_\epsilon\equiv\partial/\partial \epsilon$,
$\partial^2_\epsilon\equiv\partial^2/\partial \epsilon^2$,
$\partial_{\tau_r}\equiv\partial/\partial \tau_r$, and $\tau_r \equiv
-\sigma_T n_e(r_{\rm in}) (r-r_{\rm out})$.  When the electron density
is uniform, then the dimensionless quantity $\tau_r$ is equal to the
electron scattering optical depth in the radial direction measured
from $r_{\rm out}$. The coefficients in equations (\ref{0th3}) and
(\ref{1st}) are given in Appendix~C.

{\it Boundary Conditions}---The boundary conditions for both the
transfer equation and the system of its moments depend on the specific
problem under study. For the model problems discussed in \S5, I shall
assume that the flow is not illuminated from the outside, i.e.,
 \begin{equation}
 I^-(r_{\rm out},\theta,\epsilon)=0
 \label{inbound}
 \end{equation}
 at the outer boundary, 
and specify the radiation flux at the inner boundary as
 \begin{equation}
 I^+(r_{\rm in},\theta,\epsilon)=
 I^-(r_{\rm in},\theta,\epsilon)+4H^r_0(\epsilon)\;.
 \label{outbound}
 \end{equation}
 
For solving the moments of the radiative transfer equation I shall
specify, for all photon energies $\epsilon$, the first moment of the
monochromatic specific intensity at the inner radius, i.e.,
 \begin{equation}
 H^r(r_{\rm in},\epsilon)=H^r_0(\epsilon)
 \label{inboundmom}
 \end{equation}
as well as the ratio of the first to the zeroth moment of
the monochromatic specific intensity at the outer radius
 \begin{equation}
 \frac{H^r(r_{\rm out},\epsilon)}{J(r_{\rm
out},\epsilon)}=k(\epsilon)\;.
 \label{outboundmom}
 \end{equation}
The quantity $k(\epsilon)$ is not known a priori but is calculated
in a self-consistent way in the iterative method described below.
I shall also set at all radii 
\begin{equation}
\lim_{\epsilon\rightarrow 0} J(r,\epsilon)=
   \lim_{\epsilon\rightarrow \infty} J(r,\epsilon)  =  0
\label{Ein} 
\end{equation}
and
\begin{equation}
\lim_{\epsilon\rightarrow 0} H^r(r,\epsilon)=
   \lim_{\epsilon\rightarrow \infty} H^r(r,\epsilon)  =  0\;,
\label{Eout}
\end{equation}
which reflect the fact that the photon occupation number and the
radiation flux are finite quantities.

\section{MATHEMATICAL CHARACTER OF THE MOMENT EQUATIONS}

In the method of variable Eddington factors, the system of partial
differential equations (\ref{0th3}) and (\ref{1st}) is solved
iteratively with the radiative transfer equation~(\ref{RTEq}), given a
set of Eddington factors computed from the previous iteration. The
boundary conditions and the method used for the solution of the system
depends on the character of the partial differential operator. For
example, the system of moment equations in the frame {\em comoving\/}
with the flow (or the local Lorentz frame when gravitational effects
are taken into account) may be hyperbolic or elliptic, depending on
the velocity field (Turolla et al.\ 1995).  In this section I follow
the procedure outlined by Ames (1992, pp.~8--12) to show that the {\em
system-frame}, energy dependent moment equations always form a
parabolic system of partial differential equations, thus simplifying
the implementation of the numerical algorithm.

I define the quantities $L=\partial_\e J$ and $M^r=\partial_\e H^r$,
which together with the partial differential equations (\ref{0th3})
and (\ref{1st}), and the equations for the differentials $dJ$, $dH^r$,
$dL$, and $dM^r$, form an $8\times 8$ system of algebraic equations
for the eight derivatives
\begin{eqnarray}
\left[
\begin{array}{cccccccc}
1 & 0 & 0 & 0 & 0 & 0 & 0 & 0 \\
0 & 0 & 1  & 0 & 0 & 0 & 0 & 0  \\
0 & 0 & 0 & 1 & A_3\e^2  & 0& 0 & 0 \\
0 & f^{rr} & 0 & 0 & 0 & 0 & B_5 \e^2 & 0 \\
0 & d\tau_r  & 0 & 0 & 0 & 0 & 0 & 0 \\
0 & 0 & 0 & d\tau_r & 0 & 0 & 0 & 0  \\
0 & 0 & 0 & 0 & d\e & d\tau_r & 0 &0  \\
0 & 0 & 0 & 0 & 0 & 0 &d\e & d\tau_r  
\end{array}\right]
\cdot
\left[
\begin{array}{c}
\partial_\e J\\
\partial_{\tau_r} J\\
\partial_\e H^r\\
\partial_{\tau_r} H^r\\
\partial_\e L\\
\partial_{\tau_r} L\\
\partial_\e M^r\\
\partial_{\tau_r} M^r\\
\end{array}
\right]
&=&\qquad\qquad\qquad\qquad\nonumber
\end{eqnarray}
\begin{equation}
\qquad\qquad\qquad\qquad\qquad\qquad=
\left[
\begin{array}{c}
L\\
M^r\\
C_1-A_1 J -A_2\e L-A_4 H^r-A_5 \e M^r\\
C_2-B_1 J -B_2 \e L - B_3 H - B_4 \e M^r\\
dJ-Ld\e\\
dH^r-M^r d\e\\
dL\\
dM^r
\end{array}
\right]\;.
\end{equation}
Equating the determinant of the coefficient matrix to zero, I obtain
the characteristic equation for the system of equations,
\begin{equation}
 A_3 B_5 \e^4 \left(d\tau_r\right)^4 = 0\;.
\label{cheq}
\end{equation}
The coefficients $A_3$ and $B_5$ are proportional to the electron
temperature and the square of the bulk velocity (see Appendix~C) and
hence for the problems under consideration they are never
zero. Therefore, for a finite electron energy $\epsilon$, the system
of partial differential equations (\ref{0th3}) and (\ref{1st}) has
only one multiple, real characteristic direction, i.e., $d\tau_r=0$,
and hence is always parabolic. The difference with respect to the
moment equations in the comoving frame arises from the presence of
second order energy derivatives in the differential operator.

\section{NUMERICAL METHOD}

Equation (\ref{RTEq}) is an integro-differential equation for the
specific intensity of the radiation field and therefore any numerical
solution of this equation is not trivial to obtain. On the other hand,
the system of equations (\ref{0th3}) and (\ref{1st}) depends on two
variable Eddington factors and an unknown outer boundary condition,
which can be computed only when the solution of equation (\ref{RTEq})
for the specific intensity is obtained. Here, I employ a numerical
algorithm that is a generalization of the method of variable Eddington
factors (Mihalas 1980; see also Mihalas 1978, pp.~157--158) in order
to solve iteratively equation (\ref{RTEq}) for the specific intensity
of the radiation field and the system of partial differential
equations (\ref{0th3}) and (\ref{1st}) for its zeroth and first
moments. This algorithm has been proven to be very efficient and gives
a solution to the complete transfer equation that is limited only by
numerical accuracy, independent of the optical depth of the medium.

In this method, I first solve the moments of the transfer equation
using an initial guess for the variable Eddington factors and for the
outer boundary condition. I then compute the generalized source
function using the calculated moments of the specific intensity, solve
the radiative transfer equation, and use this solution to update the
Eddington factors. I repeat the whole procedure until convergence is
achieved.

\subsection{Solution of the Moments of the Transfer Equation}

The moments of the transfer equation depend only on radius and photon
energy, when the variable Eddington factors are known. Therefore, I
discretize the domain of solution and the differential operators of
equations (\ref{0th3}) and (\ref{1st}) on a two--dimensional mesh of
$N_\tau$ grid points over the variable $\tau_r$ and $N_{En}$ grid
points over the photon energy $\epsilon$.

The mesh of discrete grid points over the variable $\tau_r$ must
resolve the rapid change of the variable Eddington factors with
optical depth near both boundaries of the domain of solution. In order
to achieve this, I define the mesh points to be equidistant in the
quantity
\begin{equation}
Q=\left\{\begin{array}{ll}
        W\tau + (1-W) \log \tau\;,& \quad\quad \tau\le
             \frac{1}{2}\tau_m\\
        W\left(\tau_m-\tau\right)+
             (1-W)\log\left(\tau_m-\tau\right)
          & \quad\quad \tau\ge\frac{1}{2}\tau_m
\end{array}
\right.\;,
\end{equation}
where $\tau$ is the total optical depth (and not the quantity
$\tau_r$) in the radial direction measured from $r_{\rm out}$,
$\tau_m\equiv\tau(r_{in})$ is the maximum radial optical depth in the
medium, and $W$ is a parameter that allows a combination of a
logarithmic grid near the boundaries with a linear grid in the
interior. Because the spectra that emerge from Comptonizing media are
usually power laws, I use a logarithmic mesh of discrete grid points
over photon energy $\epsilon$.

The system of moment equations involve differentiation over two
density-like quantities, the radiation energy density $J$ and the
first Eddington factor $f^{rr}$, and two flux-like quantities, the
radiation energy flux $H^r$ and the second Eddington factor $g^{rr}$.
I discretize density-like quantities in a shell-centered fashion and
flux-like quantities on the grid points.  Let $F_{ij}$ be the value of
a physical quantity at the $i$--th grid point in the variable $\tau_r$
and the $j$--th grid point in photon energy. Away from the boundaries,
I differentiate density-like quantities over the variable $\tau_r$ on
the grid points as
\begin{equation}
\left(\frac{\partial F}{\partial \tau_r}\right)_{i,j}=2
\frac{F_{i+1/2,j}-F_{i-1/2,j}}{(\tau_r)_{i+1}-(\tau_r)_{i-1}}\;,
 \label{back}
\end{equation}
and flux-like quantities as
\begin{equation}
\left(\frac{\partial F_{ij}}{\partial \tau_r}\right)_{i+1/2,j}=
\frac{F_{i+1,j}-F_{i,j}}{(\tau_r)_{i+1}-(\tau_r)_i}\;,
 \label{for}
\end{equation}
in an explicit or fully implicit scheme.

In both equations I approximate the first derivative of a physical
quantity $F$ with respect to photon energy by
\begin{equation}
\left(\frac{\partial F}{\partial
\epsilon}\right)_{ij}=
\frac{F_{i,j+1}-F_{i,j-1}}{\epsilon_{j+1}-\epsilon_{j-1}}\;.
\label{1stder}
\end{equation}
and the second derivative with respect to photon energy by
\begin{equation}
\left(\frac{\partial^2 F}{\partial
\epsilon^2}\right)_{ij}=
\frac{2}{\epsilon_{j+1}-\epsilon_{j-1}}\left(
\frac{F_{i,j+1}-F_{i,j}}{\epsilon_{j+1}-\epsilon_{j}}
-\frac{F_{i,j}-F_{i,j-1}}{\epsilon_{j}-\epsilon_{j-1}}
\right)\;,
  \label{2ndder}
\end{equation}
where the index `$i$' in the quantity $F_{ij}$ may also be fractional,
corresponding to a shell center, as needed by an implicit differencing
scheme over the variable $\tau_r$.  The energy differencing
scheme~(\ref{1stder})--(\ref{2ndder}) cannot be used directly in
differencing the radiation energy density and flux in
equations~(\ref{0th3}) and (\ref{1st}) because it does not guarantee
particle conservation when only scattering is taken into account
(Chang \& Cooper 1970). This is important when the energy spectrum has
reached quasi-equlibrium and small numerical errors lead to
significant deviation from particle conservation (see eq.~[10] in
Chang \& Cooper 1970). A better differencing scheme can be deduced by
examining the properties of the differential operator at that limit.

The solution of the moment equations reaches quasi-equlibrium when
the photon mean free path is very small compared to any characteristic
length scale in the system. In that limit, $f^{rr}=1/3+{\cal O}(V)$, and
when only processes that conserve photon number are operating, the
moment equation of zeroth order becomes
\begin{equation}
\left(\partial_r+\frac{2}{r}\right)\frac{H^r}{\e}=
   n_e \sigma_{\rm T}\partial_\e \left[ VH^r+(\e-3T_e-V^2)J 
   +\left(T_e+\frac{V^2}{3}\right)\e\partial_\e J\right]\;,
  \label{cons}
\end{equation}
where $n_e$ is the dimensional electron density. The photon number
flux is simply equal to $N^r\equiv H^r/\e$ and integrating
equation~(\ref{cons}) over photon energy gives
\begin{equation}
\left(\partial_r+\frac{2}{r}\right) \int_\e N^r d\e = 0\;,
\end{equation}
i.e., photon conservation. Following Chang \& Cooper (1970), I define
a generalized photon current ${\cal F}(r,\e)$ as
\begin{equation}
{\cal F}(r,\e)=VH^r+(\e-3T_e-V^2)J
   +\left(T_e+\frac{V^2}{3}\right)\e\partial_\e J
\end{equation}
and write the zeroth and first moment equations as
\begin{eqnarray}
A_1' J + A_2' \epsilon \partial_\epsilon J + A_3'
\epsilon^2\partial^2_\epsilon J +  A_4' H^r +\partial_{\tau_r}H^r
   & = & C_1 +n_e\epsilon\partial_\epsilon {\cal F}\label{0th4}\\
B_1 J + B_2 \epsilon \partial_\epsilon J +f^{rr}\partial_{\tau_r}J+
   B_3 H^r + B_4 \epsilon \partial_\epsilon H^r +  B_5
\epsilon^2\partial^2_\epsilon H^r
   & = & C_2 \label{1st4}\;,
\end{eqnarray}
where the coefficient are given in Appendix~D. Differencing
equations~(\ref{0th4})--(\ref{1st4}) according to the
scheme~(\ref{back})--(\ref{2ndder}) results in strict photon number
conservation in the limit of small photon mean free paths.

A final point with the energy differencing is related to the
requirement that the photon energy density is everywhere
positive. This introduces a restriction on the spacing of the energy
grid, depending on the differencing scheme. For the scheme used above
it can be shown that the requirement for a positive photon energy
density results in the restriction (cf.\ eq.~[\ref{cons}] and Chang \&
Cooper 1970, eq.~[14])
\begin{equation}
\frac{\Delta\e}{\e}< \left(\frac{1}{2}\right)
  \frac{T_e+V^2/3}{\vert \e-3(T_e+V^2/3)\vert}< 
\left\{\begin{array}{ll}
        1/6\;,&\qquad \e\ll 3T_e+V^2\\
         \left(\frac{1}{2\e}\right)(T_e+V^2/3)\;, 
    & \qquad \e\gg 3T_e+V^2
\end{array}\right.\;,
\end{equation}
which can be easily met.

I solve the system of equations~(\ref{0th4}) and (\ref{1st4}) for the
zeroth and first moments of the specific intensity of the radiation
field, $J$ and $H^r$, in $N_\tau$ grid points in optical depth and
$N_{\rm En}$ grid points in photon energy. Therefore the system of
equations has $N_\tau\times N_{\rm En}$ unknowns. In all the interior
grid points ($1< i < N_\tau$; $1< j < N_{\rm En}$) I solve both
equations~(\ref{0th4}) and (\ref{1st4}) using the differencing scheme
of equations~(\ref{back})--(\ref{2ndder}). For the inner spatial
boundary, i.e., for $i=1$ and for all $j$, I solve
equations~(\ref{inboundmom}) and (\ref{1st4}), whereas for the outer
spatial boundary, i.e., for $i=N_\tau$ and for all $j$, I solve
equations~(\ref{outboundmom}) and (\ref{0th4}).  Finally, at both
energy boundaries, i.e., for all $i$ and for $j=1,N_{En}$, I set both
the zeroth and first moments of the specific intensity equal to zero,
according to boundary conditions (\ref{Ein}) and (\ref{Eout}).

The resulting system of equations is linear in both the zeroth and
first moments of the specific intensity. Defining the quantity $P_i$
as
\begin{equation}
\begin{array}{ccccc}
   P_1=J_{1,1}\;; & P_2 = H^r_{1,1}\;; & ... &
    P_{2N_\tau-1}=J_{N_\tau,1}\; 
      & P_{2N_\tau}=H^r_{N_\tau,1}\\
\end{array}\quad \mbox{etc.}
\end{equation}
the system of equations takes the form shown in Figure~1. The matrix
of coefficients is a sparse matrix and can be solved efficiently with
the biconjugate gradient method (see, e.g., Press et al.\ 1992,
pg.~209) with a preconditioner matrix. I choose the preconditioner
matrix to be the coefficient matrix, with the elements at the outer
two diagonal bands set equal to zero. The preconditioner matrix can
then be inverted efficiently with the LU-decomposition method (see,
e.g., Press et al.\ 1992, pg.~202). In this method, in order to avoid
round-off errors, I scale all the rows in the coefficient matrix so
that the highest value of the elements in each row is unity. Depending
on the complexity of the particular problem, the system of difference
equations can be solved to a fractional accuracy of $10^{-5}$ within
less than fifty iterations, which can be performed in a few seconds on
a current workstation.

\begin{figure}[t]
 \centerline{ 
\epsscale{0.7}
\plotone{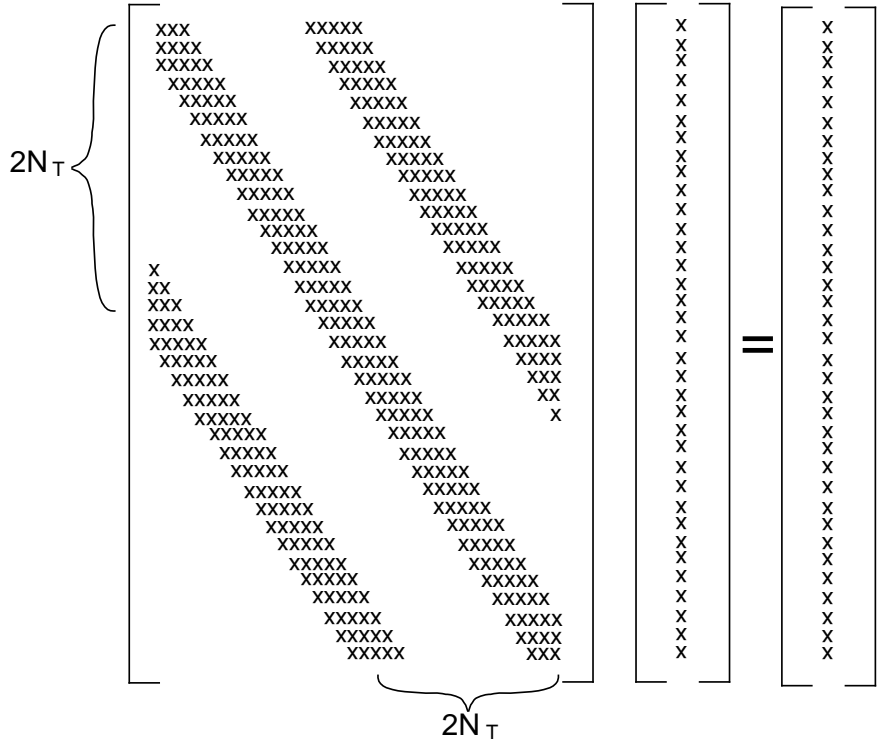}}
\figcaption[]{\footnotesize Schematic representation of the linear system of
the discretized zeroth and first moments of the
radiative transfer equation.}
\end{figure}

\subsection{Solution of the Radiative Transfer Equation}

In the iterative method of the variable Eddington factors, the
transfer equations~(\ref{RTEq}) is solved at every grid point in the
$(p-z)$ plane for the boundary conditions~(\ref{inbound}) and
(\ref{outbound}). Because the generalized source function is
calculated using the results from the previous iterations, the
solution of equation (\ref{RTEq}) is just the formal solution that can
be obtained analytically, i.e.,
\begin{eqnarray}
I^-(\tau_z,p,\epsilon) & = &
   \int_0^{\tau_z} S(t,p,\epsilon)
e^{(1-2\epsilon)(t-\tau_z)}
    dt \nonumber\\ 
I^+(\tau_z,p,\epsilon) & = & I^+(z_{\rm min},p,\epsilon)
   e^{(1-2\epsilon)(\tau_z-\tau_{zmin})}\nonumber\\
& &\qquad\qquad 
   +\int^{\tau_{zmin}}_{\tau_z} S(t,p,\epsilon)
  e^{(1-2\epsilon)(\tau_z-t)} dt\;,
  \label{formal}
\end{eqnarray}
where 
\begin{equation}
I^+(z_{\rm min},p,\epsilon) = \left\{\begin{array}{ll}
          I^-(z_{\rm min},p,\epsilon)\;, & p>r_{\rm in}\\
          I^-(z_{\rm min},p,\epsilon)+4H^r_0(\epsilon)\;, &
p< r_{\rm in}
                 \end{array}\right.
\end{equation}
and $z_{min}=0$ or $z_{min}=(r_{\rm in}^2-p^2)^{1/2}$ when $p>r_{\rm
in}$ or $p<r_{\rm in}$ respectively. In calculating the integrals in
equation~(\ref{formal}), I use a piece-wise linear interpolation of the
source function through each shell and evaluate the energy derivatives
according to equations~(\ref{1stder})--(\ref{2ndder}).

\begin{figure}[t]
 \centerline{
 \psfig{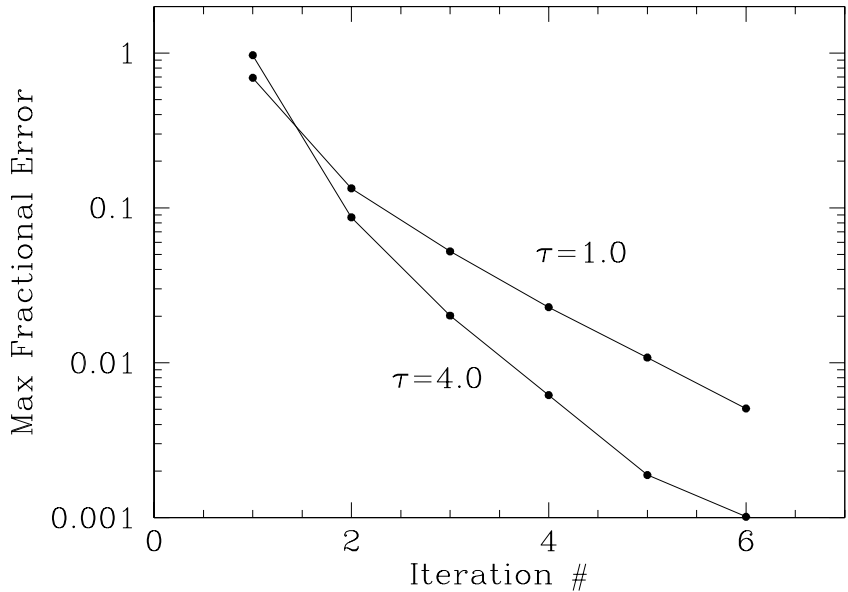}}
\figcaption[]{\footnotesize Maximum fractional error in the
   radiation energy density after each iteration, for the test
   problems discussed in \S4.3.}
\end{figure}

\subsection{Convergence and Validation}

The rapid convergence of the method of variable Eddington factors is
preserved in the current implementation for the problem of Compton
scattering. This is shown in Figure~2, where the maximum fractional
error in the radiation energy density is plotted for each iteration,
for two uniform media with different scattering optical depths. For
the model problems shown, the outer radius of the scattering medium is
set to $3 r_{\rm in}$, its electron temperature to 10~keV, and the
energy-dependence of the radiation flux at the inner boundary to be
that of a blackbody with a temperature of 0.5~keV. The converge is
typically faster for problems with higher optical depths, because I
have used $f^{rr}=1/3$ as an initial value for the first variable
Eddington factor, which is closer to the real value at the limit of
high optical depth.

The numerical method described in this section also allows for a
consistency check between the solution of the transfer equation and
the solution of the moment equations that can be performed for each
individual problem. For this test, equation~(\ref{RTEq}) can be solved
iteratively without using the solution of the system of
equations~(\ref{0th3})--(\ref{1st}), by starting with an assumed
generalized source function [e.g., $S(z,p,\epsilon)=0$], solving the
transfer equation for the specific intensity, updating the generalized
source function, and repeating until convergence is achieved. The
moments of the specific intensity can then be obtained in this way and
compared directly with the moments obtained by solving
equations~(\ref{0th3}) and (\ref{1st}).

The desired accuracy of the calculations, which is affected mainly by
the choice of the discrete mesh, depends on the particular
problem. Because the main goal of this study is the calculation of the
spectra of accreting compact objects, the X-ray colors of which are
observed to change by a few percent when the mass accretion rate
changes by a factor of $\sim 2$, I require for the solutions a
fractional accuracy of $\sim 10^{-3}$ at each grid point.

\section{RESULTS}

In this section I solve the radiative transfer equation for a number
of model problems with spherical symmetry that are related to various
astrophysical systems and in particular to media around compact
objects. In \S5.1, I discuss Comptonization of soft X-ray photons by a
static, uniform medium of hot electrons.  In \S5.2, I solve the
radiative transfer equation in a cold, divergence-free inflow;
although such a configuration requires fine tuning of the physical
conditions around an accreting object and may not occur in nature, it
is used here to provide physical understanding of the effect of
Comptonization by the bulk electron velocity. Finally, in
\S5.3, I discuss more realistic problems of spherical accretion onto
and outflows from compact stars.

\subsection{Compton Scattering in a Hot, Static Medium}

\begin{figure}[t]
 \centerline{ \psfig{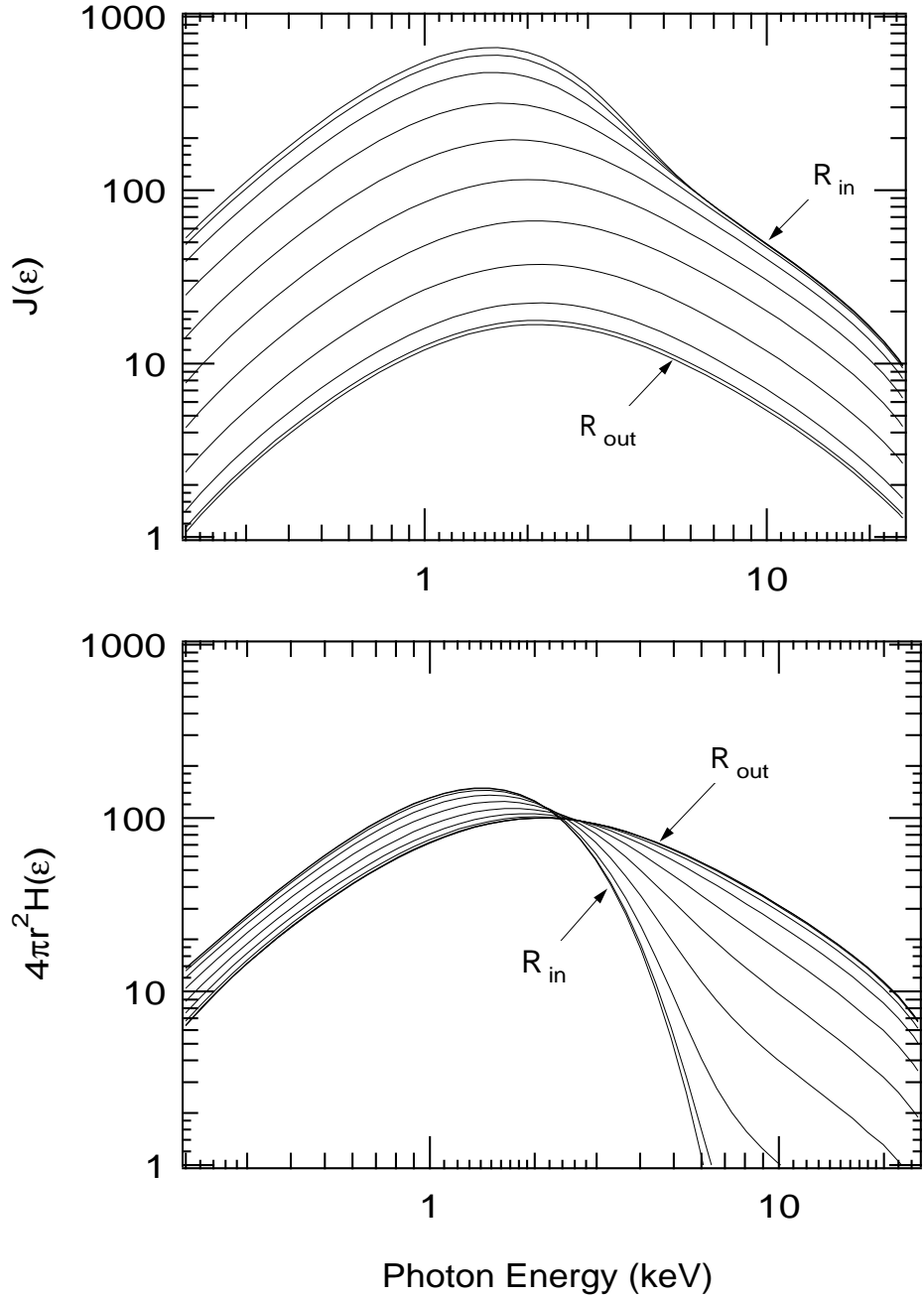}}
 \figcaption[]{\footnotesize The zeroth and first moments of the
 specific intensity of the radiation field in the static, hot medium
 discussed in \S5.1 (in arbitrary units). In both panels the curves
 correspond to radii in the medium that are equidistant in the
 parameter Q (see eq.[19]).}
\end{figure}

For the first model problem, I consider a static, spherically
symmetric, purely scattering medium and set the parameters to values
that are typical for weakly magnetic, accreting neutron stars (see
Psaltis et al. 1995). I set the outer radius of the scattering medium
to $3 r_{\rm in}$, its electron temperature to 10~keV, its electron
scattering optical depth to 4, and the energy-dependence of the
radiation flux at the inner boundary to be that of a blackbody with a
temperature of 0.5~keV; the normalization of the input flux is
arbitrary because I have assumed that there are no sources of photons
in the medium. This is a frequently solved problem that has been
analyzed in detail by, e.g., Katz (1976), Shapiro et al.\ (1976), and
Sunyaev \& Titarchuk (1980). All these authors showed that the
emerging radiation spectrum at energies much larger than the energy of
the injected photons is largely independent of the details of the
input spectrum and can be described by a power-law with an exponential
cut-off at $\sim 3 T_e$. Here I solve this simple problem to study the
fact that the variable Eddington factors are not independent of energy
even when the photon mean-free path is independent of energy and the
medium is static (see also the discussion in Paper~I).

Figure~3 shows the zeroth and first moments of the specific intensity
of the radiation field, $J$ and $H^r$, as a function of photon energy
and radius. Both quantities decrease overall with increasing radius
because of the dilution of the radiation field. The energy dependence
of the radiation energy density changes within one photon
mean-free-path from the inner boundary but remains unchanged at larger
radii. On the other hand, the radiation flux evolves throughout the
scattering medium.  This is a consequence of the fact that the system
of moment equations has been solved by imposing an inner boundary
condition on $H$ and an outer boundary condition on the ratio $H/J$
and not on $J$ itself.

\begin{figure}[t]
 \centerline{ \psfig{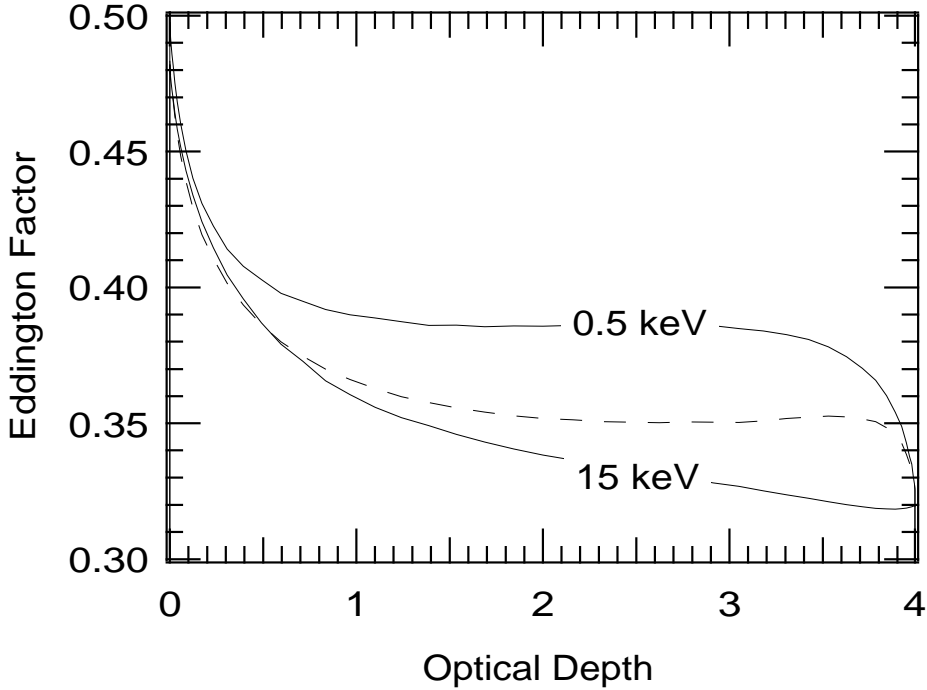}}
 \figcaption[]{\footnotesize The first Eddington factor at two photon
 energies (solid lines) as well as the ratio of the energy-integrated
 second to zeroth moments of the specific intensity of the radiation
 field (dashed line) for the model problem discuss in \S5.1.}
\end{figure}

Figure~4 shows the first Eddington factor $f^{rr}=K^{rr}/J$ at two
photon energies as well as the ratio of the energy-integrated second
to zeroth moments of the specific intensity of the radiation
field. The Eddington factor depends on photon energy even though
the scattering cross section and hence the photon mean-free-path are
mostly energy independent at these low photon energies. This is
because photons were injected into the medium with energies $\ll T_e$
and on average gain energy at each scattering. As a result, photons
emerging with low energy have experienced on average a smaller number
of scatterings than photons with higher energy and therefore their
distribution is less isotropic.

Figure~5 compares the emerging radiation spectrum obtained using the
method of variable Eddington factors with the spectra obtained using
the same boundary conditions but two energy-independent prescriptions
of the first Eddington factor that are commonly used; the second
Eddington factor does not enter the calculation when the bulk velocity
of the electrons is zero. Note that neither prescription is the
correct diffusion approximation for Compton scattering because they
both neglect the energy dependence of $f^{rr}$ on photon energy shown
in Figure~4. However, the discrepancy between the self-consistent
solution and the ones obtained using prescribed Eddington factors is
small for the prescription that depends on optical depth and can be as
large as $\sim 50$\% at high photon energies for the prescription that
is independent of optical depth. This discrepancy increases when the
ratio of the photon mean-free path to the smallest characteristic
length scale in the system increases.

\begin{figure}[t]
 \centerline{\psfig{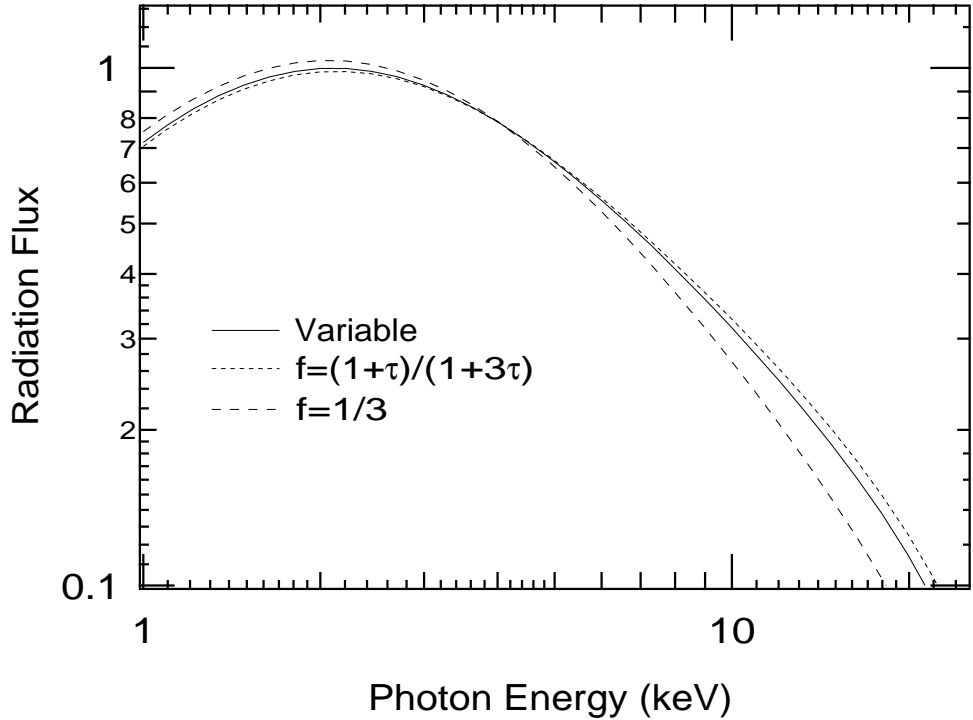}}
 \figcaption[]{\footnotesize The radiation spectrum at the outer
 boundary (in arbitrary units) obtained using the method of variable
 Eddington factors and the spectra obtained using the same boundary
 conditions but two energy-independent prescriptions of the first
 Eddington factor.}
\end{figure}

\subsection{Compton Scattering in a Divergence-Free Flow}

In a moving medium, the photons gain energy by scattering off the fast
moving electrons. As measured by an observer {\em comoving with the
medium}, the photon energy density increases mostly due to the
convergence of the flow, which is described by terms that are
proportional to $\nabla\cdot\vec{V}$ (Blandford \& Payne 1981). On the
other hand, the flux that is measured by an observer {\em at
infinity\/} increases because of the systematic energy exchange
between photons and electrons, which is described mostly by terms that
are second-order in $V$ (Psaltis \& Lamb 1997).

In order to investigate the effects of Comptonization by the bulk
electron flow and distinguish them from the effects of the convergence
of the flow, I study in this section the photon-electron interaction
and the formation of spectra in fast, divergence-free flows.  For the
model problems shown below, I set the outer radius of the scattering
medium to $30 r_{\rm in}$, the electron temperature to zero, and the
electron velocity to the divergence-free profile
$V=0.2r^{-2}$. Finally, I set the electron density to a constant
value, which arises from mass conservation in the flow, and
characterize each flow by the total electron scattering optical depth
in the radial direction.

\begin{figure}[t]
 \centerline{ \psfig{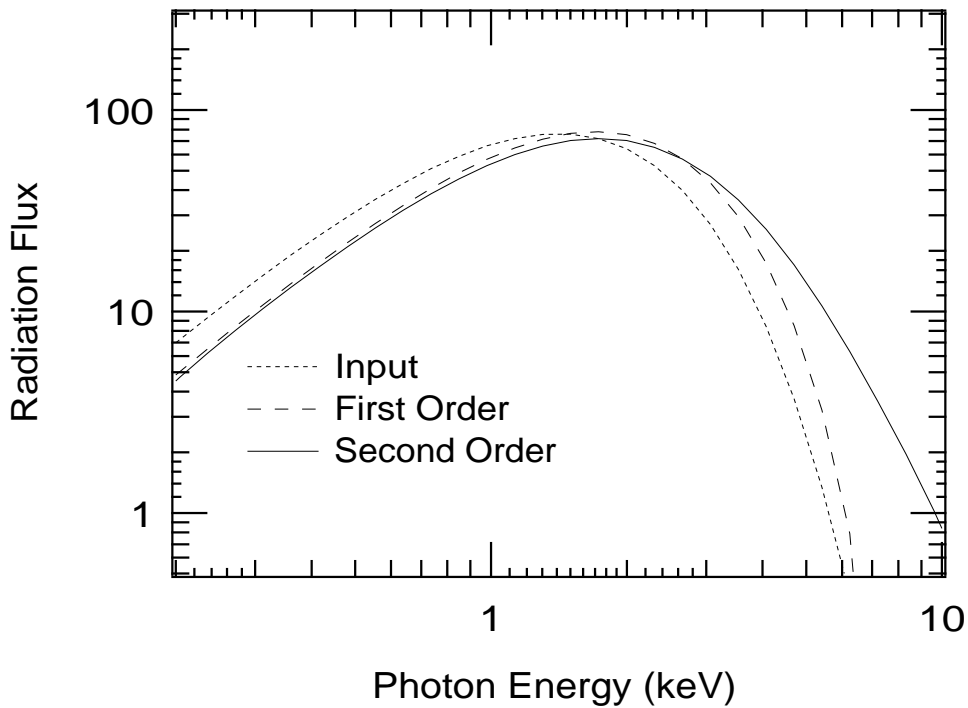}}
   \figcaption[]{\footnotesize The radiation spectrum at the inner
   boundary (dotted line), the emerging radiation spectrum when only
   terms that are first-order in the electron bulk velocity are taken
   into account (dashed line), and the emerging radiation spectrum
   that is correct to second order in the electron bulk velocity
   (solid line), in arbitrary units, for the configuration discussed
   in \S5.2.}
\end{figure}

Figure~6 shows the radiation spectrum at the inner boundary, the
emerging radiation spectrum when only terms that are first-order in
the electron bulk velocity are taken into account, as well as the
emerging radiation spectrum that is correct to second order in the
electron bulk velocity, for a flow with a scattering optical depth of
2. The terms that are first-order in the electron bulk velocity
describe mainly the effects of non-relativistic Doppler shift and
hence produce a displacement in $\log \epsilon$ of the input spectrum
towards higher photon energies (see also Psaltis \& Lamb 2000).  On
the other hand, the terms that are second-order in the electron bulk
velocity result in a systematic photon upscattering and produce a
power-law tail at photon energies higher than the average energy of
the injected photons. Note here that the approximation employed here
of keeping, in the scattering kernel, only terms that are first order
in $\epsilon$ limits the validity of the calculation to energies
$\lesssim 100$~keV.

When the total optical depth of the scattering medium, and hence the
average number of scatterings that each photon experiences, are
increased, then on average photons gain more energy by scattering off
moving electrons. As a result, the emerging radiation spectrum is
displaced towards higher photon energies, and the power-law tail
becomes flatter (see Fig.~7a). When the bulk electron velocity in the
flow increases, the average energy gain per scattering increases as
well, and hence the power-law tail also becomes flatter (Fig.~7b).  As
a result, scattering of photons by fast moving electrons produces
power-law tails at high photon energies, even in divergence-free
flows, in a way that is similar to the well understood process of
thermal Comptonization.

\begin{figure}[t]
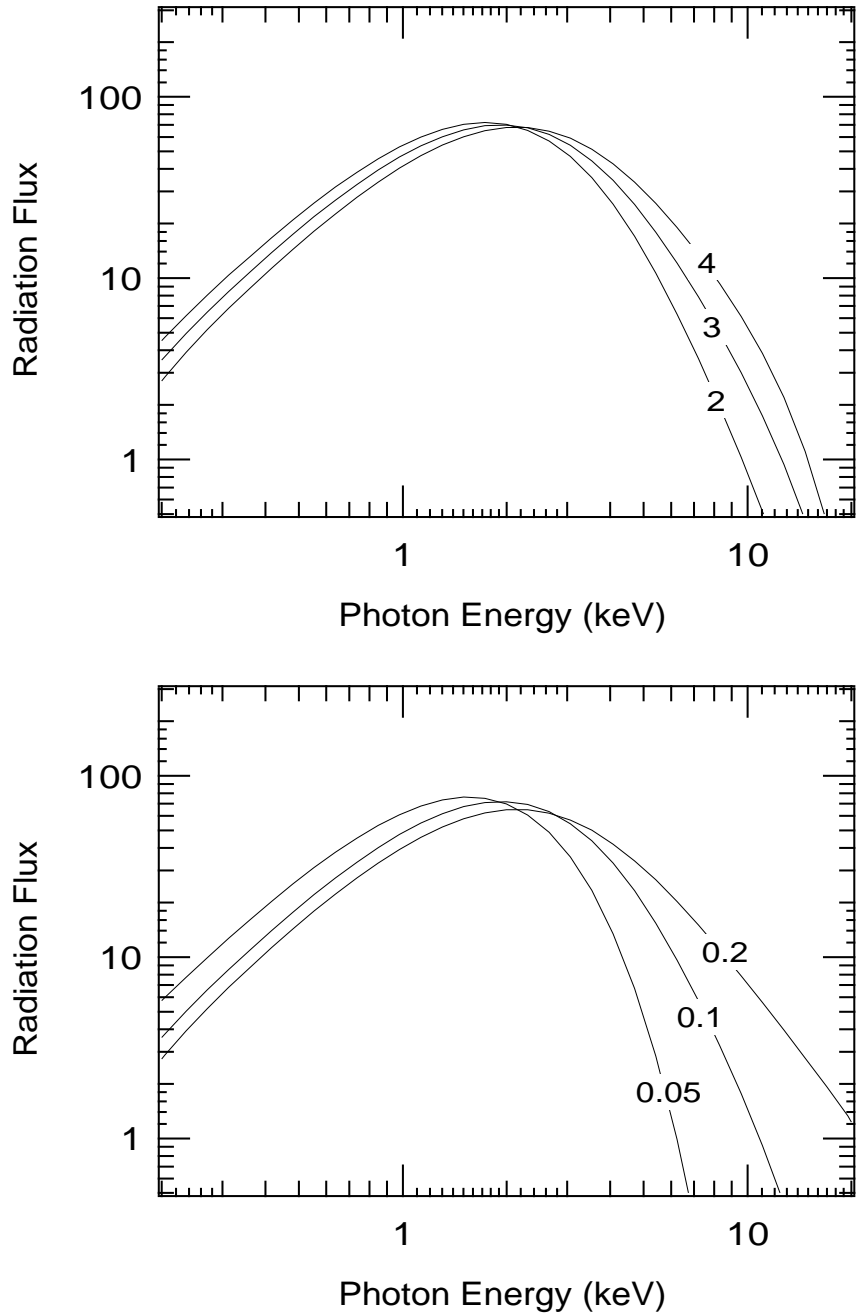

\centerline{ \psfig{file=f7a.EPSF,angle=0,height=9truecm}}
 \centerline{ \psfig{file=f7b.EPSF,angle=0,height=9truecm}}
 \figcaption[]{\footnotesize The dependence of the emerging radiation
 spectrum (in arbitrary units) on (a) the electron scattering optical
 depth and (b) the maximum electron bulk velocity, when the scattering
 optical depth is set to 5, for the configuration discussed in \S5.2.}
\end{figure}

\subsection{Compton Scattering in Inflows and Outflows}

I finally perform simple model calculations of accretion onto and
outflows from a central object by assuming a free-fall density profile
and setting everywhere in the flow the electron temperature to zero
and the inward radial velocity equal to the free-fall velocity from
infinity onto a compact object with a radius of 10~km and a
gravitational mass of $1.4 M_\odot$. I also set the outer radius of
the medium to $30 r_{\rm in}$; at these large radii the photon
mean-free path is large compared to radius and the electron bulk
velocity is small compared to the speed of light that the flow does
not effect significantly the photon distribution.  Finally, instead of
specifying the mass accretion rate onto the compact object, I specify
the total electron-scattering optical depth of the flow.

\begin{figure}[t]
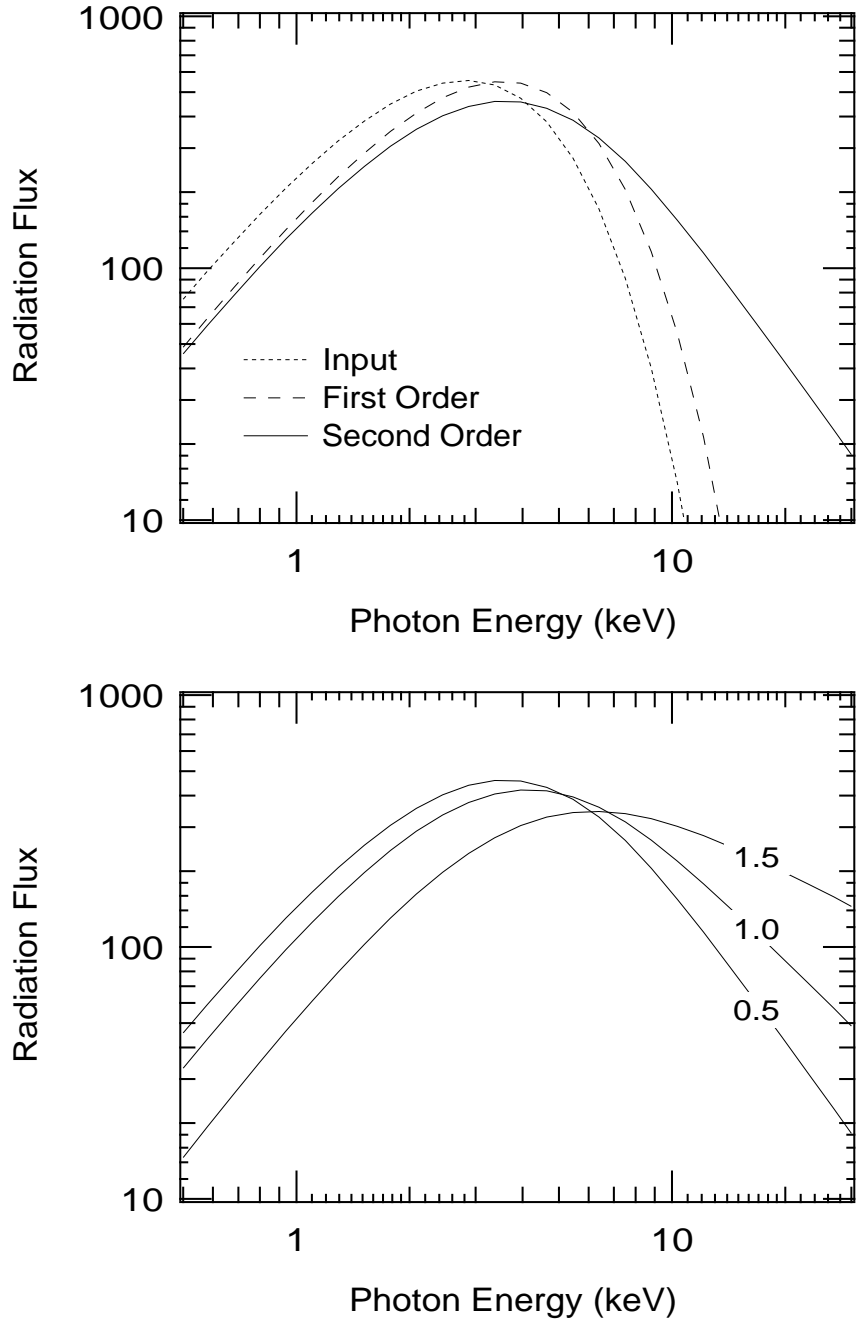

 \centerline{ \psfig{file=f8a.EPSF,angle=0,height=9truecm}}
 \centerline{ \psfig{file=f8b.EPSF,angle=0,height=9truecm}}
 \figcaption[]{\footnotesize (a) The radiation spectrum at the inner
 boundary (dotted line), the emerging radiation spectrum when only
 terms that are first-order in the electron bulk velocity are taken
 into account (dashed line), and the emerging radiation spectrum that
 is correct to second order in the electron bulk velocity (solid
 line), in arbitrary units, for the inflow problem discussed in
 \S5.3. (b) The dependence of the emerging radiation spectrum on the
 electron-scattering optical depth of the flow.}
\end{figure}

Figure~8a shows the radiation spectrum at the inner boundary, the
emerging radiation spectrum when only terms that are first-order in
the electron bulk velocity are taken into account, as well as the
emerging radiation spectrum that is correct to second order in the
electron bulk velocity. As in the model calculation discussed in
\S5.2, the terms that are first-order in the electron bulk velocity
describe mostly the non-relativistic Doppler shifts and produce a
displacement in $\log \epsilon$ of the input radiation spectrum
towards higher photon energies. On the other hand, the terms that are
second-order in the electron bulk velocity systematically upscatter
photons and produce a power-law tail at high photon energies. When the
optical depth of the flow is increased, the systematic upscattering of
photons becomes more efficient and the high-energy tail becomes
flatter.

Figure~9 shows the same information as Figure~8 but for the case of an
outflow. For this model problems I have used the same parameters as in
the inflow calculation but changed the sign of the velocity at all
radii. In an outflow, the terms that are first-order in the electron
bulk velocity produce a displacement in $\log \epsilon$ of the input
radiation spectrum towards {\it lower\/} photon energies. On the other
hand, the effect of the terms that are second-order in the electron
bulk velocity is the same as in the case of inflow, i.e., they
systematically upscatter the photons and produce a tail at high photon
energies. For the optical depths and bulk velocities considered here,
the effect of the second-order terms nearly cancels the effect of the
first-order terms and the major difference between the input and
emerging spectra is the power-law tail at high photon energies. As in
the case of the inflow, when the optical depth of the flow increases,
the power-law tail becomes flatter, but the overall effects is
significantly reduced.

\begin{figure}[t]
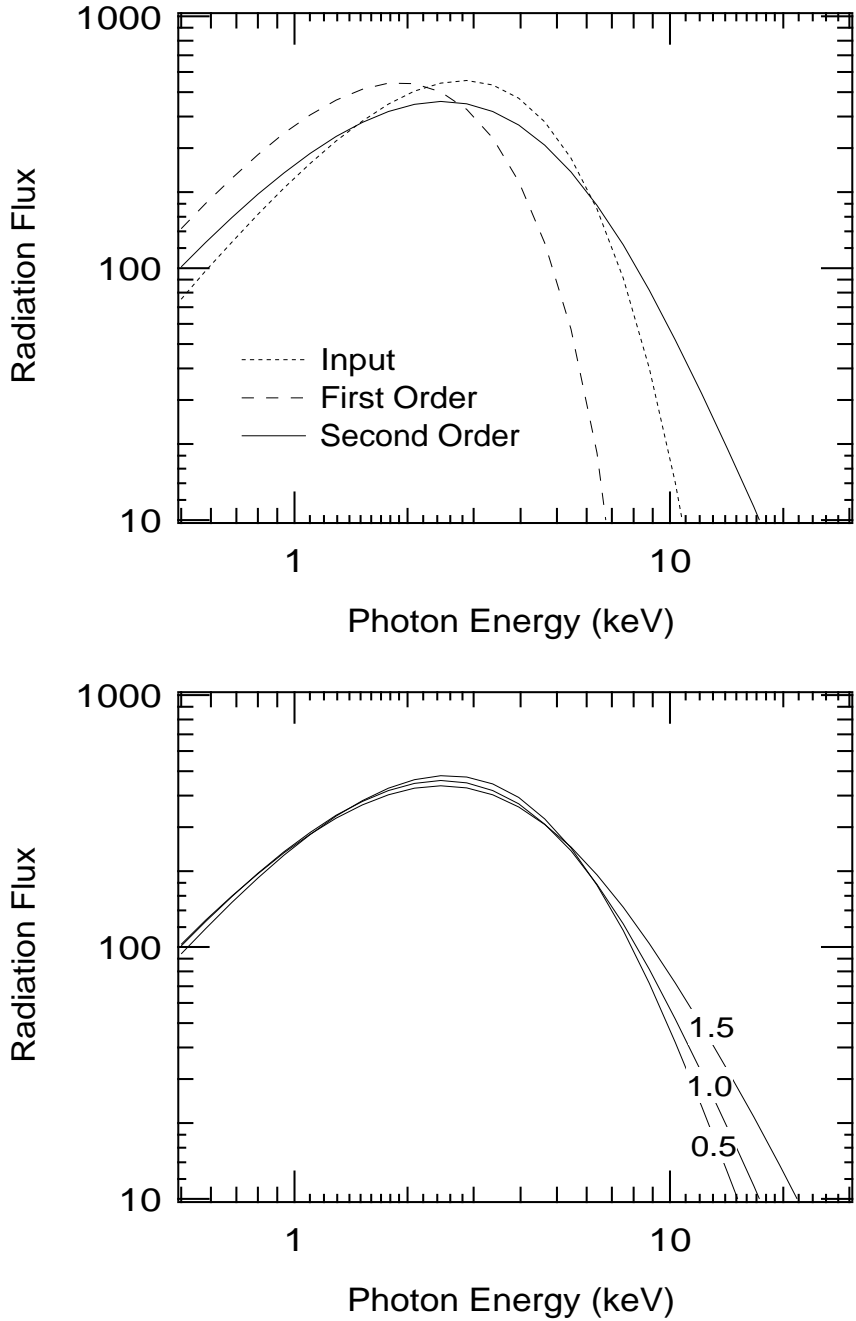

 \centerline{ \psfig{file=f9a.EPSF,angle=0,height=9truecm}}
 \centerline{ \psfig{file=f9b.EPSF,angle=0,height=9truecm}}
 \figcaption[]{\footnotesize Same as in Figure 8, but for the outflow
 problem discussed in \S5.3. }
\end{figure}

\section{DISCUSSION}

In the previous sections I described a method for solving numerically
the radiative transfer equation that describes Compton scattering in
spherically symmetric, static and moving media, and applied it to a
variety of cases related to flows around compact objects. I showed
that the first Eddington factor, which describes the degree of
isotropy of the radiation field at a given point in space, may
depend on photon energy, even though the photon mean-free-path is
energy independent, if the photons are preferably upscattered or
downscattered in energy at each scattering. I also demonstrated that
the terms that are first-order in the electron bulk velocity alter the
input spectrum in a way that is qualitatively and quantitatively
different than the effect of the terms that are of second-order.

The systematic upscattering of photons caused by bulk Comptonization
is described by the terms that are second-order in $V$. This becomes
apparent by the fact that, at each scattering with a moving electron,
the average fractional energy change of a photon in the system frame
is always positive, independent of the photon energy, and equal to
$4V^2/3$ (see, e.g., Rybicki \& Lightman 1979, ch.~7; see also the
discussion in Psaltis \& Lamb 1997). If the distribution of photon
residence times in the Comptonizing medium drops exponentially to zero
at late times, then a power-law tail is produced at high photon
energies (for total scattering optical depths $\gtrsim 1$), in analogy
to the generation of power-law tails by thermal Comptonization.  The
velocity at which the terms that are second-order in $V$ become
important depends on the ratio of the photon mean-free-path
$\lambda_{\rm mfp}$ to the shortest characteristic length scale $L$ in
the problem. In general, when $V(L/\lambda_{\rm mfp}) \sim 1$, i.e.,
in regions close to and within the photon-trapping radius, the terms
that are first- and second-order in $V$ become comparable in size (Yin
\& Miller 1995; Psaltis \& Lamb 1997a) and hence the latter should not
be neglected.

The generalization of the method to time-independent multi-dimensional
configurations is conceptually easy. However, a problem with axial
symmetry requires the solution of the radiative transfer equation in
five dimensions in phase space and the introduction of a large number of
independent variable Eddington factors. Moreover, the presence of high
flow velocities in multi-dimensions introduces sharp gradients in the
variable Eddington factors, which cannot be guessed a priori, as
required for the first iteration (Dykema, Klein, \& Castor 1996).
Even with these problems, however, the method of variable Eddington
factors will still offer a significant advantage, when compared to any
other iterative method, which arises from the fact that the Eddington
factors are bound to lie over a very narrow range of values.

The fast convergence of the numerical algorithm presented here makes
it also ideal for future extensions to transport problems coupled to
time-dependent hydrodynamics. For the case of accretion onto and
outflows from compact objects, the characteristic photon diffusion
timescale from a flow of dimension $R$ is of order
\begin{equation}
t_{\rm dif}\simeq\frac{R\tau}{c}=30\left(\frac{R}{10^6~\mbox{cm}}\right)
   \left(\frac{\tau}{1}\right)~\mu\mbox{s}\;,
\end{equation}
which is significantly smaller than the corresponding $t_{\rm dyn}\sim
1$~ms dynamical timescale. As a result, for most cases the structure
of the radiation field can be obtained by solving the time-independent
radiative transfer problem for each snapshot of the fluid properties.
Given that the variable Eddington factors will be only marginally
different between successive timesteps, the convergence of the
algorithm will be even faster than what is shown in Figure~2.

\acknowledgments

I am grateful to Fred Lamb for his support and numerous useful
discussions, to Dimitri Mihalas for his help on solving numerically
radiative transport equations, and to D.\ Swesty for bringing to my
attention the biconjugate gradient method for solving linear systems
of equations. It is also a pleasure to thank Luca Zampieri and Feryal
\"Ozel for many helpful discussions and comments and an anonymous referee
for a very careful review of the manuscript. I acknowledge the support
of a postdoctoral fellowship of the Smithsonian Institution. This
research was also supported in part by NSF grant AST~93-15133 and NASA
grant NAG~5-2925.

\newpage 

\appendix

\begin{center}
{\bf Appendices}
\end{center}

\section{Moments of the Specific Intensity in Spherical
Geometry}

In defining the moments of the specific intensity in the
system frame I shall use the quantities
\begin{equation}
u(z,p,\e)\equiv \frac{1}{2}\left[I^+(z,p,\e)
   +I^-(z,p,\e)\right]
\end{equation}
and
\begin{equation}
v(z,p,\e)\equiv \frac{1}{2}\left[I^+(z,p,\e)
   -I^-(z,p,\e)\right]\;.
\end{equation}
The zeroth moment of the specific intensity is then
\begin{equation}
J(r,\e) = \frac{1}{r}\int_0^r u(z,p,\e) dz\;,
\end{equation}
where $r$ depends on $z$ and $p$ through the relation
$r^2=z^2+p^2$. Because of the assumed spherical symmetry, the
only non-zero component of the first moment of the specific
intensity is the $r$--component:
\begin{equation}
H^r(r,\e) = \frac{1}{r^2}
   \int_0^r z v(z,p,\e) dz\;.
\end{equation}
Similarly, the non-zero components of the second, third, and
fourth moments of the specific intensity are
\begin{eqnarray}
K^{rr}(r,\e) & = & \frac{1}{r^3}
   \int_0^r z^2 u(z,p,\e) dz\nonumber\\
K^{\theta\theta}(r,\e) = K^{\phi\phi}(r,\e) 
   & = & \frac{1}{2}\left(J-K^{rr}\right)\;,\\
Q^{rrr}(r,\e) & = & \frac{1}{r^4}
   \int_0^r z^3 v(z,p,\e) dz\nonumber\\
Q^{\theta\theta r}(r,\e) = Q^{\theta r
   \theta}(r,\e) = Q^{r \theta\theta}(r,\e)
   & = & \frac{1}{2}\left(H^r-Q^{rrr}\right)\nonumber\\
Q^{\phi\phi r}(r,\e) = Q^{\phi r
   \phi}(r,\e) = Q^{r \phi\phi}(r,\e)
   & = & \frac{1}{2}\left(H^r-Q^{rrr}\right)\;,\nonumber\\
R^{rrrr}(r,\e) & = & \frac{1}{r^5}
   \int_0^r z^4 u(z,p,\e) dz\nonumber\\
R^{\theta\theta\theta\theta} (r,\e) & = &
  \frac{3}{8}\left(J-2 K^{rr}+R^{rrrr}\right)\nonumber\\
R^{\phi\phi\phi\phi} (r,\e) & = &
  \frac{3}{8}\left(J-2 K^{rr}+R^{rrrr}\right)\nonumber\\
R^{rr\theta\theta}(r,\e)=R^{r\theta r\theta}(r,\e)=
   ... & = & \frac{1}{2}\left(K^{rr}-R^{rrrr}\right)
  \nonumber\\
R^{rr\phi\phi}(r,\e)=R^{r\phi r\phi}(r,\e)=
   ... & = & \frac{1}{2}\left(K^{rr}-R^{rrrr}\right)
  \nonumber\\
R^{\theta\theta\phi\phi}(r,\e)=
  R^{\theta\phi\theta\phi}(r,\e) = ...
 & = & \frac{3}{8}R^{rrrr}\;.
\end{eqnarray}
Finally, I shall also use the first two Eddington
factors defined by
\begin{eqnarray}
f^{rr} & \equiv & \frac{K^{rr}}{J}\\
g^{rr} & \equiv & \frac{Q^{rrr}}{H^r}\;.
\end{eqnarray}

\section{The Generalized Source Function}

Using equation (A8)--(A12) of Paper I, I derive the
generalized source function for a spherically symmetric
system:
\begin{equation}
S^\pm(z,p,\e)\equiv S_1^\pm
   +\e S_2^\pm
   +T_e S_3^\pm + V S_4^\pm + V^2 S_5^\pm\;,
\end{equation}
where
\begin{eqnarray}
S_1^\pm& = & \frac{3}{8}\left[ 3 - \cos^2\theta +
   \left(3 \cos^2\theta - 1\right) f^{rr}\right] J\\
S_2^\pm& = & \frac{3}{8} \left\{
    \left(-1 + \e\partial_\e\right)
   \left[ 3 - \cos^2\theta+\left(3\cos^2\theta-1\right)f^{rr}
         \right]\right. J
\nonumber\\
& &\quad\quad+\left.\left(1-\e\partial_\e\right)
    \left[5-3\cos^2\theta+\left(5\cos^2\theta-3\right)
         g^{rr}\right]H^r\cos\theta\right\}\\
S_3^\pm& = & \frac{3}{4}\left\{
   \left[3\cos^2\theta-1-
    3\left(3\cos^2\theta-1\right)f^{rr}\right]
   J\right.\nonumber\\ 
& & \quad\quad+ 2\left[1-3g^{rr}+
   \left(5 g^{rr}-3\right)\cos^2\theta \right]H^r
   \cos\theta\nonumber\\ 
& & \quad\quad+
 \frac{1}{2}\left(-2\e \partial_\e
   +\e^2\partial^2_\e\right)
   \left[3-\cos^2\theta+\left(3\cos^2\theta-1\right)f^{rr}
   \right]J\nonumber\\
& & \quad\quad+ \frac{1}{2}\left(2\e \partial_\e
   -\e^2\partial^2_\e\right)
   \left[5-3\cos^2\theta+\left(5\cos^2\theta-3\right)g^{rr}
   \right]H^r\cos\theta\\
S_4^\pm & = &  V\cos\theta I +\frac{3}{8}V\left\{
   4\cos\theta\left[2-\cos^2\theta
      +\left(3\cos^2\theta-2\right)f^{rr}\right]J
    \right.\nonumber\\
 & &\quad\quad
  +\left[-1-5\cos^2\theta+\left(3\cos^2\theta-
       1\right)g^{rr}\right]H^r\nonumber\\
 & & \quad\quad
  -\cos\theta\partial_\e \left[3-\cos^2\theta+
    \left(3\cos^2\theta-1\right)f^{rr}\right]J\nonumber\\
 & & \quad\quad\left. + \partial_\e
  \left[\frac{1}{2}\left(3-\cos^2\theta\right)+
   \left(3\cos^2\theta-1\right)g^{rr}\right]H^{r}\right\}\\
S_5^\pm & = & \frac{3}{4}V^2 \left\{ \left[
 -5\cos^4\theta+\frac{21}{2}\cos^2\theta-1
  -2\cos^2\theta\left(2-\cos^2\theta\right)
 \e\partial_\e \right.\right.\nonumber\\
& & \quad\quad\quad\quad
 \left.+\frac{1}{4}\cos^2\theta\left(3-\cos^2\theta\right)
  \e^2 \partial_\e^2\right]J\nonumber\\
& & \quad\quad+\left[4-10\cos^2\theta+
   \left(3+\cos^2\theta\right)\e \partial_\e
  -\frac{1}{2}\left(3-\cos^2\theta\right)
  \e^2\partial_\e^2\right]\cos\theta H\nonumber\\
& & \quad\quad +\left[
  \frac{1}{2}\left(30\cos^4\theta-37\cos^2\theta+7\right)
  +2\left(1-3\cos^4\theta\right)\e\partial_\e
 \right.\nonumber\\
& & \quad\quad\quad\quad \left.
 +\frac{1}{4}\left(3\cos^4\theta-3\cos^2\theta+4\right)
  \e^2\partial_\e^2\right]f^{rr}K^{rr}\nonumber\\
& & \quad\quad +\left[
 \left(6\cos^2\theta-4\right)+\left(3\cos^2\theta-3\right)
 \e\partial_\e
 -\frac{1}{2}\left(3\cos^2\theta-1\right)\e^2
 \partial_\e^2\right]\cos\theta g^{rr} H^r
 \nonumber\\
& & \quad\quad \left.
  +\left(2\cos^2\theta-1\right)\left( 1+2\e
  \partial_\e +\frac{1}{2}\e^2\partial_\e^2
  \right)R^{rrrr}\right\}
\end{eqnarray}
where $cos\theta\equiv z/r$ and the variable Eddington
factors and the moments of the specific intensity are
defined as in Appendix A.

\section{Coefficients of the Moment Equations}

Using equations (34) and (40) of Paper I, I derive
\begin{eqnarray}
A_1 & = & n_e \left[\e + V^2 \left(\frac{1}{10}\e \partial_\e
   +\frac{11}{20}\e^2\partial^2_\e\right)f^{rr}\right]
   \nonumber\\
& & \quad\quad-\left[1+\frac{1}{2}V^2 
   +\left(f^{rr}+\frac{1}{2}\right)V^2 \e\partial_\e 
   + \frac{1}{2}f^{rr}V^2 \e^2\partial^2_\e\right]\chi\\
A_2 & = & n_e \left\{\e-2T_e-\frac{2}{3}V^2
   + \left[\frac{1}{10}\left(f^{rr}-\frac{1}{3}\right)
      + \frac{22}{20}\e\partial_\e f^{rr}\right]V^2\right\}\\
A_3 & = & n_e \left[T_e+\frac{1}{3}V^2 +
   \frac{11}{20}\left(f^{rr}-\frac{1}{3}\right)V^2\right]\\
A_4 & = & -\frac{2}{\tau_0-\tau_{\rm r}}+V\chi + V \e\partial_\e
     \chi\\ 
A_5 & = & n_e V\\
C_1 & = & -\chi S_e +\frac{1}{6}V^2 S_e 
   \left(\e\partial_\e - \e^2 \partial^2_\e\right)\chi
   +\frac{1}{6}V^2 \e \left(\partial_\e S_e\right)
   \left(1-2\e\partial_\e\right)
   \chi\nonumber\\
& & - \frac{1}{6} V^2 \e^2 \left(\partial^2_\e S_e\right) \chi
\end{eqnarray}
and
\begin{eqnarray}
B_1 & = & \frac{1}{10} n_e V \left(9 f^{rr}-7 -\e\partial_\e
   f^{rr} \right)+V
   f^{rr}\left(1+\e\partial_\e\right)\chi\nonumber\\ 
& & \quad\quad +\partial_{\tau_r}f^{rr}-\frac{3}{\tau_0-\tau_{\rm r}}
   \left(f^{rr}-\frac{1}{3}\right)\\
B_2 & = & -\frac{1}{10}n_e V \left(f^{rr}+3\right)\\
B_3 & = & n_e \left[ -1 +\frac{12}{5}\e-\frac{2}{5}T_e
   -\frac{1}{2}V^2+\frac{1}{10}V^2\left(-1-3\e\partial_\e
   -\partial^2_\e\right)g^{rr}\right]\nonumber\\
& & \quad\quad 
   -\left[1+\frac{1}{2}V^2+V^2\left(g^{rr}
   +\frac{1}{2}\right) \e\partial_\e
   +\frac{1}{2}V^2g^{rr}\e^2\partial^2_\e\right]\chi\\
B_4 & = & n_e \left[-\frac{2}{5}\e+\frac{4}{5}T_e
   +\frac{9}{10}V^2
   -\frac{3}{10}V^2\left(1+\frac{2}{3}\e\partial_\e\right)
     g^{rr}\right]\\
B_5 & = & n_e\left[-\frac{2}{5}T_e -\frac{1}{10}V^2
   \left(g^{rr}+3\right)\right]\\
C_2 & = & -\frac{2}{3}V\chi S_e+
   \frac{1}{3}V \chi \e\partial_\e S_e +
   \frac{1}{3}V S_e \e\partial_\e \chi\;,
\end{eqnarray}
where $\tau_0\equiv\tau_{\rm r}(r=r_{\rm in})$ and I have suppressed the
dependence of the various quantities on spatial position and
photon energy.

\section{Coefficients for Differencing the Moment Equations}

Using equations (34) and (40) of Paper I I derive
\begin{eqnarray}
A_1' & = & n_e V^2 \left(\frac{1}{10}\e \partial_\e
   +\frac{11}{20}\e^2\partial^2_\e\right)f^{rr}
   \nonumber\\
& & \quad\quad-\left[1+\frac{1}{2}V^2 
   +\left(f^{rr}+\frac{1}{2}\right)V^2 \e\partial_\e 
   + \frac{1}{2}f^{rr}V^2 \e^2\partial^2_\e\right]\chi\\
A_2' & = & n_e \left[\frac{1}{10}\left(f^{rr}-\frac{1}{3}\right)
      + \frac{22}{20}\e\partial_\e f^{rr}\right]V^2\\
A_3' & = & n_e 
   \frac{11}{20}\left(f^{rr}-\frac{1}{3}\right)V^2\\
A_4' & = & -\frac{2}{\tau_0-\tau_{\rm r}}+V\chi + V \e\partial_\e
     \chi\\ 
C_1' & = & -\chi S_e +\frac{1}{6}V^2 S_e 
   \left(\e\partial_\e - \e^2 \partial^2_\e\right)\chi
   +\frac{1}{6}V^2 \e \left(\partial_\e S_e\right)
   \left(1-2\e\partial_\e\right)
   \chi\nonumber\\
& & - \frac{1}{6} V^2 \e^2 \left(\partial^2_\e S_e\right) \chi\;,
\end{eqnarray}
where $\tau_0\equiv\tau_{\rm r}(r=r_{\rm in})$ and I have suppressed
the dependence of the various quantities on spatial position and
photon energy.

\end{document}